\documentclass[%
 reprint,
 amsmath,amssymb,
 aps,
]{revtex4-1}

\usepackage{graphicx}
\usepackage{dcolumn}
\usepackage{bm}

\usepackage{hyperref}
\hypersetup{
	colorlinks = true,
	linkcolor = black,
	filecolor = black,
	urlcolor = black,
	citecolor = black
}

\usepackage{verbatim}

\begin{document}

\preprint{APS/123-QED}

\title{Extension of $\nu$MSM model and possible explanations of recent astronomical and collider observations}

\author{Krishna Kulkarni}
 \email{kdkulkarni93@iitkgp.ac.in; kdkulkarni93@gmail.com}
\affiliation{%
 Fifth year integrated Master's student,\\
 Department of Physics,\\
 Indian Institute of Technology Kharagpur, Kharagpur 721302, India
}%


\date{\today}

\begin{abstract}
Here I present the extension to ${\nu}$MSM model by adding a $U(1)'$ gauge symmetry under which right-handed fermions including sterile neutrinos and exotic Higgs scalar are charged. This model explains 3.5 keV line observed by XMM-Newton telescope as well as Galactic Center Excess(GCE) detected by Fermi-LAT satellite. The proposed model also accounts for recent diphoton excess of mass around 750 GeV produced in ATLAS and CMS detectors. Cosmological inconsistencies such as baryon asymmetry can also be explained. Verification of presented model is possible with current experimental techniques and limitations. Careful validation and analysis of the model needs to be done in future.

\end{abstract}

\pacs{Valid PACS appear here}
\maketitle


\section{Introduction}

Extension of Standard Model(SM) has been studied widely since it's birth. Many theories have been developed to extend SM in various ways with variety of motivations. Extension of SM in neutrino sector is one of the good ways for extension since it's motivation lies behind previously unexplained experimental observations. Since we now know that left-handed SM neutrinos are massive, it is natural from seesaw mechanism to postulate the existence of $N_{I}$, right-handed sterile neutrinos (${\nu}$MSM model)\cite{nuMSM}.\\

In this paper, I extend $\nu$MSM model ($\mathcal{N} = 3$) by adding a $U(1)'$ gauge symmetry under which right handed fermions including sterile neutrinos, and exotic Higgs scalar are charged. This model gives rise to interesting explanations of different unexplained observations and phenomena. Section II discusses the model and put constraints on masses of right-handed sterile neutrinos $N_{I}$. Section III is devoted to explanation of 3.5 KeV line observed by XMM-Newton telescope. Section IV explains constraints in $\nu$MSM model and need for it's extension. It also presents explanation of Galactic Center Excess(GCE) detected by Fermi-LAT satellite. Section V explains diphoton excess of mass around 750 Gev produced in ATLAS and CMS detectors. In Section VI, possible methodology for verification of model in current experimental limit is discussed. Main points of the model are summarized in Section VII.

\section{Model}

In this section, I introduce the Lagrangian obtained from adding $U(1)'$ gauge symmetry to $\nu$MSM model. Adding another $U(1)'$ symmetry to Standard Model Higgs ($\phi$) gives massless gauge boson. Therefore to make new gauge boson ($E_{\mu}$) obtained from $U(1)'$ symmetry massive, there is a need for a new scalar $\Phi$ which can get the vacuum expectation value. Here, $E_{\mu}$ gets mass from same exotic Higg's sector $\Phi$, which can also be an indication of new TeV scale physics. Similar type of extension is presented in this reference\cite{TSRAY}. Also it is natural while extending SM to think of three right-handed sterile neutrinos $\mathcal{N} = 3$. In section IV it is explained that minimum number of right-handed sterile neutrinos needed to explain dark matter is three.
\begin{equation}
\mathcal{L} = \mathcal{L}' - {\frac{1}{4}}E_{{\mu}{\nu}}{E^{{\mu}{\nu}}} + {\frac{1}{2}}{{m_{E}}^{2}}E_{\mu}E^{\mu} + {g_{N}}{E_{\mu}}{\bar{f}}{{\gamma}^{\mu}}{P_{R}}f
\end{equation}

$$\mathcal{L}' = \mathcal{L}_{SM} + {\bar{N}_{I}}i{D_{\mu}}{{\gamma}_{\mu}}N_I - F'_{{\alpha}I}{\bar{L}_{\alpha}}{\phi}{N_{I}} - \frac{M_I}{2}{{\bar{N}_{I}}^{c}}N_I$$\\

where, ${{D}_\mu} = {\partial_{\mu}} + i{g_{N}}{{E}_{\mu}(x)}$, $\phi$ is Standard Model Higgs, $N_{I} = \{N_{1},N_{2},N_{3}\}$ and term with $f$ is right-handed fermion couplings with $E_{\mu}$.\\

$\nu$MSM model explains the light mass of neutrinos compared with other fermions with the condition that right-handed sterile neutrinos must be much heavier (seesaw mechanism). We keep this condition in our analysis while extending the model and put constraints on the masses of three right-handed sterile neutrinos as : $m_{N_{1}} \sim \mathcal{O}(7 KeV)$ which can explain 3.5 keV line\cite{LymanAlpha, XRay1,boyarsky}. Taking masses of other two sterile neutrinos  $m_{N_{2}} \approx m_{N_{3}} \sim \mathcal{O}(5-60 GeV)$ which are degenerate can account for explanation of baryon asymmetry\cite{boyarsky_Shaposhnikov} as well as Galactic Center Excess(GCE). (Original proposal of baryogenesis in singlet-fermion oscillations is presented in this reference\cite{baryogenesis_1}. See Refs.\cite{baryogenesis_2,nuMSM} for detailed quantitative discussions on baryogenesis.) It is shown in following sections that from above constraints on the masses of right-handed sterile neutrinos all the observations and phenomena mentioned can be explained.

\section{Explanation of 3.5 keV line observed by XMM-Newton telescope}

Recently, 7 keV dark matter candidate has received much attention due to recent observation of unidentified line at 3.25 $\pm$ 0.02 keV X-ray spectra of M31 galaxy and Perseus galaxy cluster observed by MOS\cite{MOS} and PN\cite{PN} cameras of XMM-Newton telescope. The case is made stronger by observation and analysis of 73 galaxy clusters\cite{73_galaxy}. Furthermore if only {$\nu$}MSM model  with $\mathcal{N} = 3$ is considered, then bounds on mass of dark matter particle are $m_{DM} \approx (1-10) keV$. Where, the lower bound comes from CMB and matter power spectrum inferred from Lyman-$\alpha$ forest data\cite{LymanAlpha} and upper bound is limited by X-ray observations\cite{XRay1}. Clearly particle $N_{1}$ satisfies this constraint. If interaction of $N_{1}$ to SM particles is only through mixing angle, then it is found that\cite{DW} :

\begin{equation}
{\tau}_{DM} = 7.2\times10^{29} sec \left[\frac{10^{-8}}{\sin^{2}(2\theta)}\right]{\left[\frac{1 keV}{m_{DM}}\right]^{5}}
\end{equation}

After careful analysis of 3.5 keV line, obtained mass of particle is $m_{DM} = 7.06 \pm 0.06 keV$ and mixing angle is $sin^{2}(2{\theta}) = (2-20)\times10^{-11}$\cite{boyarsky}. Which is in good agreement with experimental bounds till now.

\vspace{0.7cm}
\includegraphics[scale = 0.19]{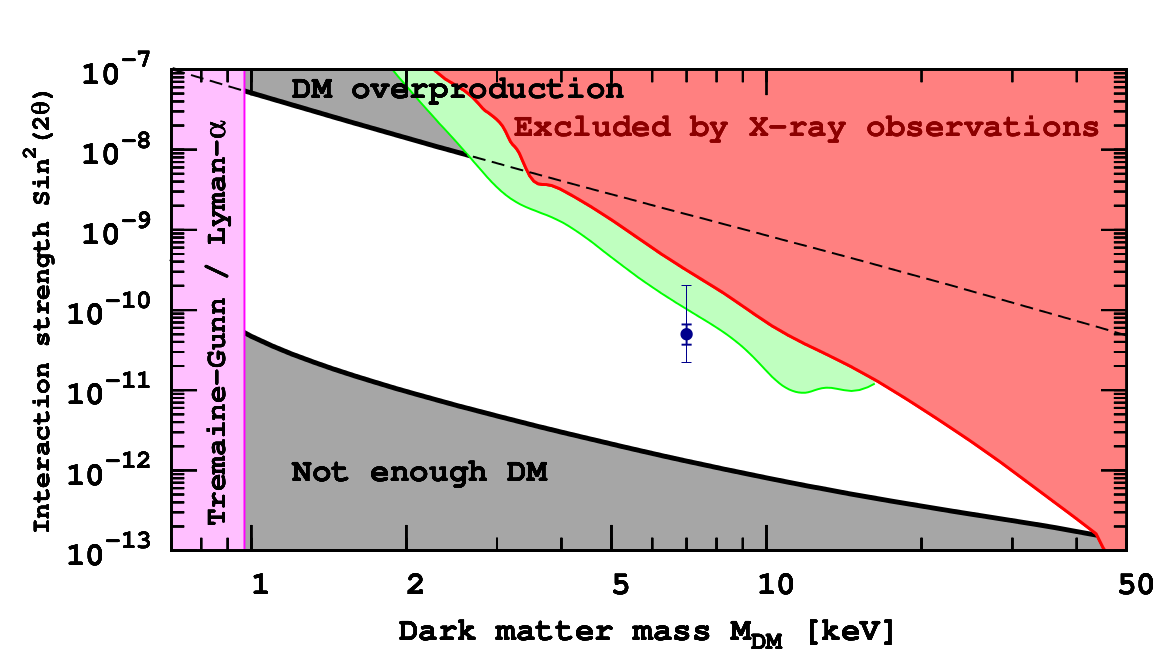}
FIG. 1 : (Image Reference : \cite{boyarsky}) Constraints on sterile neutrino dark matter within $\nu$MSM\cite{DMfig_1}. Recent upper bounds on the mass from\cite{DMfig_2,DMfig_3} are shown in green. Similar to older bounds are marked by red. At masses below ∼ 1 keV dwarf galaxies would not form and that will be a lower bound\cite{DMfig_4,DMfig_5}. Blue point corresponds to the best-fit value from M31 if the line comes from DM decay. Thick error bars are for $\pm 1\sigma$ limits on flux. And thin error bars correspond to uncertainty in the dark matter distribution in the center of M31.
\vspace{0.7cm}

In FIG. 1, points corresponding to white region are those where sterile neutrino constitute 100$\%$ of dark matter and their properties agree with the existing bounds. Within the gray regions too much (or not enough) dark matter would be produced in a minimal model like $\nu$MSM.\\

Similar type of analysis is also presented in Ref.\cite{73_galaxy} where signal from 73 galaxy clusters is analyzed. Even though these two are independent analyses, both the results are consistent with each other and the model presented in this paper.

\section{Explanation of Galactic Center Excess(GCE) detected by Fermi-LAT satellite}

Intriguing excess of $\gamma$-rays has been observed at the galactic center and has triggered large number of studies\cite{GCE1,GCE2,GCE4}. This specially extended excess of $\sim$ 1-3 GeV gamma rays from region surrounding galactic center is very well fit by 5-60 GeV particle annihilating into $b\bar{b}$ (for 20-60 GeV particle) or $\tau\bar{\tau}$ (for 5-20 GeV particle), with annihilation cross section $\left\langle{\sigma}v\right\rangle = (1-3)\times10^{-26}cm^{3}/s$\cite{GCE3,GCE5}. Even though the other possibilities of origin of GCE are there, such as through collisions of high energy protons accelerated by galaxy's supermassive black hole with gas, the possibility of dark matter origin can not be denied.\\

Now let's discuss the need to extend ${\nu}$MSM model. If only $\nu$MSM model is considered, then model have to have $\mathcal{N} \geq 3$ to account for dark matter and to explain neutrino oscillations, and if $\mathcal{N} = 3$, then model can have at most one dark matter candidate\cite{nuMSM_constraint}, which in this case is $N_{1}$. Hence ${\nu}$MSM has exhausted the possibility of other dark matter candidate and as shown below, adding one $U(1)'$ gauge symmetry will provide new mechanism for GCE explanation.\\

\vspace{0.7cm}
\includegraphics[width=\columnwidth]{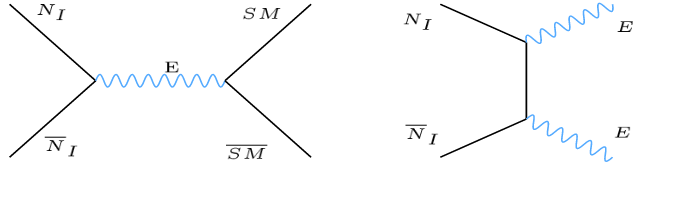}
FIG. 2 : (a) Left : s-channel $m_{N_{I}} < m_{E}$, (b) Right : t/u-channel $m_{N_{I}} > m_{E}$
\vspace{0.7cm}

For one possible explanation of GCE, let's analyze the possible implications of Lagrangian $\mathcal{L}$. One of the strengths of this model is that it can provide a new physics scale of $\mathcal{O}(1-10TeV)$ through vacuum expectation value of new exotic scalar. If we want this new $E_{\mu}$ field coupling strength, $g_{N}$ to be similar to weak coupling, then since $m_{E} \sim g_{N} \left\langle{\phi}\right\rangle$, it is obvious that $m_{E} \sim \mathcal{O}(100 GeV)$. Since $m_{N_{2}} \approx m_{N_{3}} \sim \mathcal{O}(10-60 GeV)$ and both $N_{2}$ and $N_{3}$ are charged under new $U(1)'$ gauge symmetry, both can decay into $E_{\mu}$ but from the possible interactions shown in FIG. 2, only s-channel survive. $E_{\mu}$ can decay into $b\bar{b}$ or $\tau\bar{\tau}$ which can account for GCE. Since both $N_{2}$ and $N_{3}$ have degenerate masses and similar interactions, it is very difficult to separate out signals from individual particle. Also, $N_{1}$ can be easily excluded from this type of interaction due to very light mass. More detailed analysis on similar type of extended model with Dirac fermions is presented in Ref.\cite{TSRAY}.\\

\section{Explanation of diphoton excess produced in ATLAS and CMS detectors}

Recent excess of diphoton around 750GeV is reported by ATLAS\cite{diphoton_ATLAS} as well as CMS\cite{diphoton_CMS} detector. Although the signals are 3.6$\sigma$ and 2.6$\sigma$ in respective experiments, it would indicate the discovery of new physics at TeV scale if this significance increases in future. Landau-Yang\cite{Landau,Yang} theorem says that only spin-0 or spin-2 particle can decay into two photons. Assuming spin-0 particle it is a strong indicative of electroweak hierarchy problem.\\

The extended model presented here naturally get an exotic scalar $\Phi$ which provides mass to $E_{\mu}$. Since in Section IV it is shown that vacuum expectation value of $\Phi$ can be  of $\mathcal{O}(1-10TeV)$. If we choose the value of the quartic coupling $\lambda \sim \mathcal{O}(0.02-0.2)$ (which is consistent with our intuition), mass of the $\Phi$ will be around $m_{\Phi} \sim 750GeV$.\\

Now, considering $\Phi$ as a source of diphoton signal, since no other decay channel is observed in both CMS and ATLAS other than diphoton, there should be significant branching fraction into photons. This implies that effective coupling with photon must be considerably large. For example, one of the mechanisms discussed recently\cite{Csaki} assumes only sizable coupling of $\Phi$ with Standard Model particles is to photons via operator :

$$\frac{c_{{\gamma}{\gamma}}}{v}{\Phi}F^{2}$$

Which gives near resonance cross section at 8 TeV and 13 TeV for spin-0 resonance with mass 750 GeV as :

$${\sigma}_{8 TeV} \approx 31 fb \left(\frac{\Gamma}{45 GeV}\right)Br^{2}(\Phi \rightarrow \gamma\gamma)$$
$${\sigma}_{13 TeV} \approx 162 fb \left(\frac{\Gamma}{45 GeV}\right)Br^{2}(\Phi \rightarrow \gamma\gamma)$$

Successful explanation of diphoton excess requires relatively large partial photon width $\Gamma_{{\gamma}{\gamma}} \sim 15 GeV$ which implies $c_{{\gamma}{\gamma}} \sim 0.16$. For detailed calculations, please refer to\cite{Csaki} where it is shown that proton scattering with two photon fusion can provide significant cross section to produce the 750 GeV resonance without gluon couplings.\\

\section{Possible methodology for verification of model}

Till now the model has explained existence of all $N_1$, $N_2$ and $N_3$. Now let's focus our attention to verification of model in current experimental limits. As explained in Section IV, that $\nu$MSM can have at most one dark matter candidate. Hence $N_2$ and $N_3$ can not act as a dark matter candidates and they can have shorter lifetime. Decrease in lifetime of these particles can induce higher mixing with left-handed SM neutrinos. Possible mechanism for detecting such mixing can be proposed which can provide a strong evidence for existence of this model in nature.\\

Many parameters of the models are constrained by observations mentioned in previous sections as well as from cosmological considerations and experiments on neutrino masses and oscillations. In order for this model to be verified, one of the active neutrinos must be very light $m_{{\nu}1} \sim \mathcal{O}(< 10^{-5}eV)$\cite{nuMSM_constraint}, which fixes the masses of other two active neutrinos as $m_{{\nu}2} \approx 9\times10^{-6}eV$ and $m_{{\nu}3} \approx 5\times10^{-2}eV$ for normal hierarchy or else $m_{{\nu}2,{\nu}3} \approx 5\times10^{-2}eV$ for inverted hierarchy. Hence, effective Majorana mass for neutrinoless double beta decay can be determined.\\

If the diphoton excess detected in CMS and ATLAS is confirmed with new scalar $\Phi$ with mass $m_{\Phi} \sim 750 GeV$ then it will strengthen the argument provided by this model and be a leading step in verification of model. Also, it can pave new ways to verify and constraint the model.

\section{Summary}

The model presented in this paper looks elegant and promising due to it's ability to explain different mutually exclusive observations as well as theoretical inconsistencies in SM and cosmology given below :

\begin{itemize}

\item 3.5 keV line observed by XMM-Newton telescope
\item Galactic Center Excess(GCE) detected by Fermi-LAT satellite
\item Diphoton excess of mass around 750 GeV produced in ATLAS and CMS detectors
\item Baryon asymmetry
\item Small mass of neutrinos through seesaw mechanism
\end{itemize}

Although this model explains the things mentioned above, it may not be correct. The full theory may be much more complicated. Work presented in this paper is mostly argumentative and provides general idea of the model. Quantitative as well as experimental work needs to be done in order to improve the constraints of the model as well as it's verification.\\

{\bf{Acknowledgements}}

This work is done as a part of winter project at Physical Research Laboratory(PRL), Ahmedabad, India. I would like to thank my project guide \emph{Prof. Subhendra Mohanty} for guidance and for many wonderful discussions with him that led to many ideas introduced in this document. I also thank PRL staff for providing excellent facilities and services during my stay.

\bibliography{main}

\end{document}